\documentclass[vci-paper]{vciart}
\usepackage{amssymb}

\begin{document}

\begin{frontmatter}

\title{Design and Status of IceCube}


\author{Martin Kestel for the IceCube collaboration\thanksref{fullList}}
\thanks[fullList]{See \cite{JAAhrensEtAl2003} for a full author list}
\ead{mka@phys.psu.edu}

\address{104 Davey Lab, PMB 29, Penn State University, University 
Park, PA 16802}

\begin{abstract}
IceCube is a kilometer-scale high energy neutrino detector that builds
on the wealth of experience accumulated with its smaller predecessor,
AMANDA. An international collaboration has begun construction of key
components of the IceCube detector and  deployment operations at the
South Pole will begin in late 2004.

The underlying design of the IceCube detector and of the DAQ system
are presented here, emphasizing the digital optical modules (DOMs) as
the smallest discrete IceCube building block. The event reconstruction
critically relies on a relative timing accuracy from DOM to DOM of a
few nanoseconds over inter-DOM separations of up to
1$\,$km$\,$. 
\end{abstract}

\end{frontmatter}

\section{Physics goals}\label{physicsGoals}

Through the detection of very high-energy neutrinos (threshold a few
100$\,$GeV), IceCube \cite{IceCubeHomePage,IceCubeIcrc2003}
will open a new window on the universe. By viewing astronomical
sources with neutrinos as astronomical messengers, it will address
fundamental questions in high energyastrophysics, particle physics and
cosmology. Through the detection of surface electrons and muons, the
associated IceTop surface array will allow us to study the chemical
composition of high energy cosmic rays ($E \sim 10^{18}$\,eV) and will
also help calibrate IceCube and provide a background veto. IceCube and
underwater neutrino telescopes \cite{UnderwaterHomePages} share
scientific interests, such as searches for steady or variable neutrino
emission from point like source candidates like active galactic nuclei
(AGN), supernova remnants (SNR), microquasars and gamma ray bursts
(GRB). By virtue of the low ambient noise level in the ice, the
ability to detect low-energy supernova neutrinos as an increase in the
overall trigger rate is unique to IceCube among all UHE neutrino
detectors. On the more speculative side, searches for neutrinos from
annihiliations of weakly interacting massive particles (WIMPs), for
magnetic monopoles and other exotic particles like strange quark
matter or SUSY Q-balls can be listed (see
e.g. \cite{SpieringICRC1242,HalzenAstroPh0311004}). 

\section{Detector design and status}\label{detectorDesignAndStatus}

IceCube will consist of 4800 digital optical modules (DOMs), organized in 
80 strings, each with 60 DOMs attached, buried in the ice at depths of 
1450$\,$m to 2450$\,$m$\,$. DOMs will have a vertical spacing of
17$\,$m and the strings will be regularly spaced horizontally 
by 125$\,$m$\,$. At each string location two IceTop tanks, each
containing two DOMs frozen in ice, will be deployed. The buried DOMs
will have an effective surface area of around 1$\,$km$^2\,$, promising
optimal sensitivity for neutrinos in the energy range of 1 to
10000$\,$TeV while being able to trigger on all higher- and on some of
the lower-energy neutrinos, including MeV-bursts
\cite{iceCubeConfigurationStudy}. The positions of IceCube strings,
and the IceTop tanks deployed above them, are shown in Figure 1. 
Simulations have shown that IceCube's sensitivity to possible signals
is roughly constant for a wide range of feasible configurations.

Digital optical modules (DOMs) form the fundamental building blocks of
the IceCube detector. Each DOM contains a 10$\,$'' Hamamatsu R-7081
photo multiplier (PMT). The high voltage for the PMT is converted in
the DOM from its 48$\,$V DC power supply to achieve the design gain of
around 5$\cdot10^7\,$. Within a DOM, the PMT signal is split into two
copies, with one used for triggering and the other delayed and then
digitized if the threshold condition is met. Digitization occurs in
two types of DOM-resident digitizers, to extend the digitization time
while keeping the resolution high at early times. There is a set of
two four-channel ASIC analog transient waveform digitizers (ATWDs) and
a commercial 40$\,$MHz FADC with up to 256 samples and 16 bit
resolution available. The two ATWDs operate at 300$\,$MHz and at none,
16, 32, 64 or 128 sample depths with 8- or 16-bit resolution. The two
ATWDs are fed signals in a ping-pong manner, reducing DOM dead time to
less than 1$\,$\%$\,$. The first three channels of each of the ATWDs
are fed signals that have been amplified with factors of 16, 4 and
2/3. This combination of ATWDs and FADC ensures the design dynamic
range of up to 200 photo electrons (p.$\,$e.) within the first
15$\,$ns and up to 2000 p.$\,$e. within the first 5$\,\rm{\mu
  s}\,$. The fourth ATWD channel can be connected to various inputs
like the DOM-clock ticks or LED driving currents, creating a versatile
diagnosis and calibration tool. The DOMs further contain a 405$\,$nm
LED flasher board, producing programmable light flashes of various
intensities detectable by other modules in the array. This capability
is useful for studying ice properties and calibrating the relative
positions of DOMs. A preliminary version of a DOM, deployed in an
IceTop tank in January 2004, is shown in Figure 2. 

The DOM mainboard has a free-running timer which needs to be
synchronized with nanosecond accuracy to GPS time, requiring
re-calibration roughly every minute. As shown in Figure 3,
a surface circuit sends a bipolar signal at a GPS-latched time
$t_1\,$, received at a time $t_2\,$. After a certain, fixed time
interval $\delta_t\,$, an identical circuit in the DOM sends an
identical bipolar pulse  to the surface, detected at a time
$t_4\,$. The cable transmission time is then:  $t_{Down} = t_{Up} =
(t_4 - t_1 - \delta_t) / 2\,$. This calibration reduces signal time
spread to the inevitable contribution from light scattering in the
Antarctic ice. 

Currently a fully digital, TCP/IP-based approach for the DAQ system is
under development, following closely the modular structure of the
experimental setup: 
each {\it String Processor} stores DOM-data and
passes trigger primitives on to the {\it InIce Trigger}, which, after
examining trigger primitives from all {\it String Processors}, sends
its trigger decisions to the {\it Global Trigger}. The {\it Global
  Trigger} combines {\it InIce Trigger}, {\it IceTop Trigger} and
other (external) information to form its decision. If positive, the
{\it Event Builder} is instructed to retrieve DOM data from the 
{\it String Processors} and assembles them to IceCube events that get
passed to the {\it Online Filter Cluster} for further processing. All
of these DAQ system elements are implemented in commercial computers.

The drilling process has been improved in several aspects compared to
the AMANDA procedure: setup time for a season will be only three to five
weeks, 60$\,$cm diameter holes will be drilled with water of
90$^\circ\,$C$\,$ from a number of heaters with a total power of
5$\,$MW (vs. 2$\,$MW for AMANDA), a larger hose diameter reduced drill
time to 40$\,$h$\,$, and the fuel consumption will be lowered by about
30$\,$\%$\,$. With an estimated string drop time of around
20$\,$h$\,$, it should be possible to deploy 16 or more strings per
austral summer season, leading to a construction time of five to six
years for the entire detector. 

\section{Summary}\label{summary}

With the assembly and testing of the first batch of
DOMs under way, the IceCube collaboration is on track for deployment
of the first set of strings at the end of 2004. The digital approach to
readout and triggering, together with the sophisticated time
calibration, will help to overcome the challenges posed by the sheer
size of the detector and the time spreads induced by the Antarctic ice
as a detector medium, enabling IceCube to produce useful data for
scientific purposes after just the first few deployments.

\section{Acknowledgements}\label{acknowledgements}

This research was supported by the following agencies: National
Science Foundation -- Office of Polar Programs, National
Science Foundation -- Physics Division, University of Wisconsin Alumni
Research Foundation, USA; Swedish Research Council, Swedish Polar
Research Secretariat, Knut and Alice Wallenberg Foundation, Sweden;
German Ministry for Education and Research, Deutsche
Forschungsgemeinschaft (DFG), Germany; Fund for Scientific Research
(FNRS-FWO), Flanders Institute to encourage scientific and
technological research in industry (IWT), Belgian Federal Office for
Scientific, Technical and Cultural affairs (OSTC), Belgium; Inamori
Science Foundation, Japan; FPVI, Venezuela; The Netherlands
Organization for Scientific Research (NWO).

\section{Figure captions}

Fig. 1: Aerial view of South Pole and positions of the 
IceTop tanks resp. the IceCube strings (black), Spase-2 stations
(grey, dense, regular foreground pattern \cite{SPASE2EASARRAY}) and
the AMANDA strings (larger grey pattern). Courtesy
V. Papitashvili. \label{South-Pole-And-IceCube} 

Fig. 2: The first DOM frozen into a prototype IceTop tank at South
  Pole (Jan. 2004). Photo by John Kelley / NSF. \label{domPicture}

Fig. 3: DOM time calibration, see text. \label{timingCalibrationFigure}


\begin{thebibliography}{99}

\bibitem{JAAhrensEtAl2003} J.A. Ahrens et al., Astroparticle Physics
  20 (2004) 507 

\bibitem{IceCubeHomePage} IceCube Homepage http://icecube.wisc.edu

\bibitem{IceCubeIcrc2003} S.Yoshida for the IceCube Collaboration,
  Proc. 28th ICRC (2003) 1369 

\bibitem{UnderwaterHomePages}
Antares http://antares.in2p3.fr/ Baikal
http://www-zeuthen.desy.de/baikal/baikalhome.html Nemo
http://nemoweb.lns.infn.it/ Nestor http://www.nestor.org.gr/ 

\bibitem{SpieringICRC1242} C. Spiering for the Amanda collaboration,
  Proc. 27th ICRC (2001) 1242 

\bibitem{HalzenAstroPh0311004} F.Halzen astro-ph/0311004

\bibitem{SPASE2EASARRAY} http://ast.leeds.ac.uk/haverah/spase2.shtml

\bibitem{iceCubeConfigurationStudy} M. Leuthold, Proc. Workshop on
  Large Neutrino Telescopes, Zeuthen 1998,
  http://www.ifh.de/nuastro/publications/conferences/proc.shtml 

\end{thebibliography}
\end{document}